\documentclass[prl,aps,showpacs,floatfix,twocolumn,superscriptaddress,nofootinbib]{revtex4-1}
\usepackage{amsmath}
\usepackage{amsfonts}
\usepackage{amssymb}
\usepackage{revsymb}
\usepackage{graphicx}
\usepackage{booktabs}

\newcommand{\eref}[1]{Eq.~(\ref{#1})}
\newcommand{\tref}[1]{Table~\ref{#1}}
\newcommand{\fref}[1]{Fig.~\ref{#1}}

\def\vop#1{\mathbf{\hat{#1}}}

\def\ket#1{|\,#1 \,\rangle}
\def\bra#1{\langle \, #1 \,|}

\usepackage{pdfpages}
\usepackage{etoolbox} 

\makeatletter
\patchcmd{\@outputpage@head}{\@ifx{\LS@rot\@undefined}{}{\LS@rot}}{}{}{}
\makeatother

\begin{document}
\title{Precision measurements on the $^{138}$Ba$^+$ $6s\;{}^2S_{1/2} - 5d\;{}^2D_{5/2}$ clock transition}
\author{K. J. Arnold}
\email{cqtkja@nus.edu.sg}
\affiliation{Centre for Quantum Technologies, National University of Singapore, 3 Science Drive 2, 117543 Singapore}
\affiliation{Temasek Laboratories, National University of Singapore, 5A Engineering Drive 1, 117411 Singapore}
\author{R. Kaewuam}
\affiliation{Centre for Quantum Technologies, National University of Singapore, 3 Science Drive 2, 117543 Singapore}
\author{S. R. Chanu}
\affiliation{Centre for Quantum Technologies, National University of Singapore, 3 Science Drive 2, 117543 Singapore}
\author{T. R. Tan}
\affiliation{Centre for Quantum Technologies, National University of Singapore, 3 Science Drive 2, 117543 Singapore}
\affiliation{Department of Physics, National University of Singapore, 2 Science Drive 3, 117551 Singapore}
\author{Zhiqiang Zhang}
\affiliation{Centre for Quantum Technologies, National University of Singapore, 3 Science Drive 2, 117543 Singapore}
\author{M. D. Barrett}
\email{phybmd@nus.edu.sg}
\affiliation{Centre for Quantum Technologies, National University of Singapore, 3 Science Drive 2, 117543 Singapore}
\affiliation{Department of Physics, National University of Singapore, 2 Science Drive 3, 117551 Singapore}
\date{\today}

\begin{abstract}
Measurement of the $^{138}$Ba$^+$ ${}^2S_{1/2} - {}^2D_{5/2}$ clock transition frequency and $D_{5/2}$ Land\'e $g_J$ factor are reported.  The clock transition frequency $\nu_{\mathrm{Ba}^+}=170\,126\,432\,449\,333.31\pm(0.39)_\mathrm{stat}\pm(0.29)_\mathrm{sys}\,$Hz, is obtained with accuracy limited by the frequency calibration of the maser used as a reference oscillator.  The Land\'{e} $g_J$-factor for the ${}^2D_{5/2}$ level is determined to be $g_{D}=1.200\,367\,39(24)$, which is a 30-fold improvement on previous measurements.  The $g$-factor measurements are corrected for an ac-magnetic field from trap-drive-induced currents in the electrodes, and data taken over a range of magnetic fields underscores the importance of accounting for this systematic.
\end{abstract}
\maketitle
Singly-ionized Barium has been considered for a variety of applications in atomic physics including parity non-conservation (PNC) investigations \cite{fortson1993possibility}, optical clocks \cite{sherman2005progress}, and quantum information \cite{dietrich2010hyperfine, inlek2017multispecies, hucul2017spectroscopy}.  Being a monovalent atom, it is also amenable to precise theoretical predictions of atomic properties \cite{iskrenova2008theoretical,safronova2010relativistic}, making it a testbed for theoretical and experimental techniques.  This has led to high accuracy measurement of a variety of atomic properties including Land{\'e} $g$-factors~\cite{marx1998precise,knoll1996experimental,hoffman2013radio,lewty2013experimental}, matrix elements \cite{woods2010dipole,sherman2008measurement}, branching ratios from the $P_{1/2}$ and $P_{3/2}$ levels \cite{kurz2008measurement, de2015precision, dutta2016exacting}, lifetime measurements of the lower lying metastable $D$-states \cite{yu1997radiative,auchter2014measurement}, fine-structure splittings \cite{whitford1994absolute} and hyperfine structure of odd isotopes \cite{blatt1982precision,lewty2012spectroscopy}.

The $^{138}$Ba$^+$ $S_{1/2} - D_{5/2}$ clock transition shares many of the advantages of the corresponding transitions in $^{88}$Sr$^+$ and $^{40}$Ca$^+$.  In addition, the upper-state lifetime of $\sim30\,\mathrm{s}$ would allow long interrogation times.  Moreover, dynamic-decoupling, as recently demonstrated in Sr$^+$~\cite{shaniv2018quadrupole},  is directly applicable to Ba$^+$ and opens the door to multi-ion clock implementations~\cite{aharon2019robust}.  Our objective in this study is three-fold: (i) determination of the clock transition frequency for which an accurate value has not yet been reported, (ii) an improved accuracy measurement of the $D_{5/2}$ $g_J$-factor, and (iii) validation of Ba$^+$ as a trap diagnostic tool for characterizing rf-induced ac-magnetic fields. The ac-magnetic field assessment is relevant for subsequent Lu$^+$ clock experiments in the same apparatus.

In this work, the clock frequency is measured with sub-Hertz level accuracy, which is accomplished by averaging over six Zeeman transitions to eliminate dominant linear Zeeman shifts and tensor shifts.  From a combination of four transitions, the ratio of Zeeman splittings between the $S_{1/2}$ and $D_{5/2}$ levels is extracted.  The $S_{1/2}$ Land\'e $g$-factor, $g_S$, is known to the $10^{-8}$ level \cite{marx1998precise}, so this ratio provides an accurate determination of the $D_{5/2}$ $g$-factor, $g_D$.  As ac-magnetic fields arising from rf-currents in the electrodes can compromise the accuracy of this determination, an Autler-Townes splitting on the clock transition is used to accurately characterize this systematic~\cite{gan2018oscillating}.  

\begin{figure}
\includegraphics[width=\columnwidth]{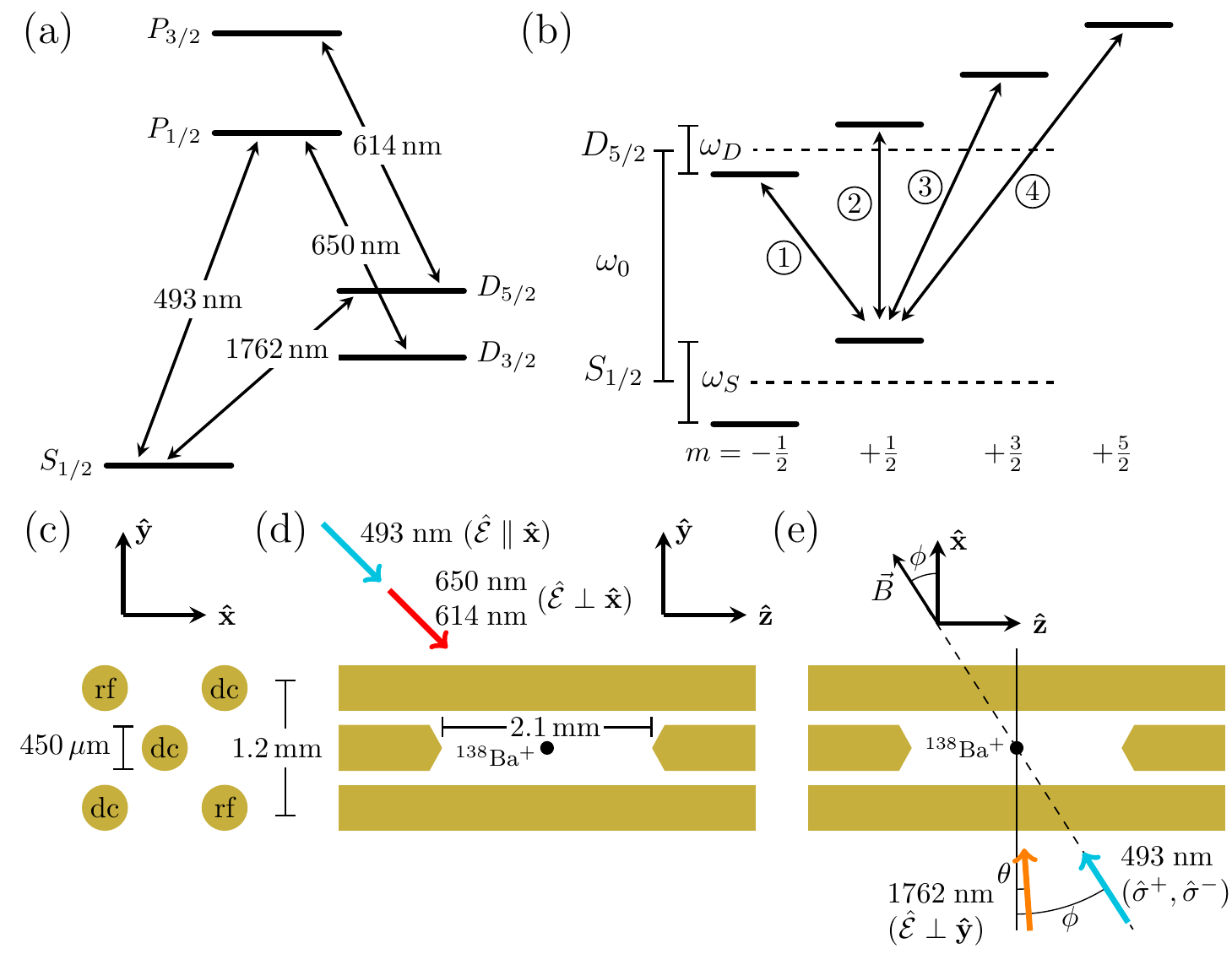}
\caption{(a) Low-lying level structure of $^{138}$Ba$^+$ showing transitions relevant to this work. (b) Optical clock transitions used in the this work. The absolute frequency of the $S_{1/2} - D_{5/2}$ transition, $\omega_0$, is obtained from the average of transitions labeled 2-4 and their Zeeman-symmetric counterparts. The ratio of the Zeeman splittings, $\omega_D/\omega_S$, is from transitions 1-2 and their Zeeman-symmetric counterparts. (c-e) End, side, and top views of the ion trap electrodes.  Polarizations and geometric orientations of lasers indicated. The magnetic field $\vec{B}$ is oriented at angle $\phi = 33(2)^\circ$ for the absolute frequency measurement, and $\phi = 0$ for the $g_D$ measurement.}
\label{levels}
\end{figure}

The experiments are carried out using a four-rod linear Paul trap with axial end-caps described in previous work~\cite{arnold2018blackbody,arnold2019oscillating}. The rf potential at a frequency near $\Omega_\mathrm{rf}=2\pi\times 20.7\,\mathrm{MHz}$ is delivered via a quarter-wave helical resonator. Together with static potentials of 5 V on the end-caps and -0.5 V on two of the diagonally opposite rods, the measured trap frequencies for a single $^{138}$Ba$^+$ ion are $\sim (890,770,240)\,\mathrm{kHz}$, with the lowest frequency along the trap axis, defined as $\vop{z}$.  The quantization axis is defined by a magnetic field applied in the $xz$-plane at angle $\phi$ with respect to $\vop{x}$ (\fref{levels}e). For the absolute frequency and $g_D$ measurements, $\phi$ is set to 33(2)$^\circ$ and 0$^\circ$, respectively. 

The level structure of $^{138}$Ba$^+$ is shown in \fref{levels}a along with the relevant transitions used in this work.  Doppler cooling is provided by driving the 493- and 650-nm transitions, where $\approx2.7\%$ of the fluorescence at 650\,nm is detected on a single photon counting module (SPCM) for state determination~\cite{arnold2019measurements}.  The $D_{5/2}$ level is populated by driving the $S_{1/2}-D_{5/2}$ clock transition at 1762\,nm and depopulated by driving the $D_{5/2}-P_{3/2}$ transition at 614\,nm.  Cooling and repumping light at 493, 614, and 650\,nm are all collinear and propagate at 45 degrees to the $\vop{z}$-axis.  For $\phi= 0$, the 493-nm cooling beam is $\pi$-polarized, whereas the 614, and 650-nm beams are polarized perpendicular to the magnetic field ($>0.2\,$mT) to avoid dark states in the $D$-levels.  State preparation into the $m=\pm1/2$ states of the $S_{1/2}$ levels is facilitated by two additional 493-nm beams that are polarized $\sigma^+$ and $\sigma^-$. 

Spectroscopy of the clock transition is implemented using an external-cavity-diode laser (ECDL), which is phase locked to an optical frequency comb (OFC).  The short term ($<10\,$s) stability of the OFC is derived from a $\sim 1\,\mathrm{Hz}$ linewidth laser at 848 nm which is referenced to a 10\,cm long ultra-low expansion (ULE) cavity with finesse of $\sim4\times10^5$. For longer times ($\gtrsim 10\,\mathrm{s}$), the OFC is steered to an active hydrogen maser (HM) reference. The frequency of the maser is calibrated to the SI (International System of Units) second by continuous comparison to a GPS timebase~\cite{endrun} in combination with the circular-T reports from the International Bureau of Weights and Measures  (BIPM)~\cite{dube2017absolute,hachisu2017absolute}. 

For the absolute frequency measurements, the 1762\,nm laser is continuously on and intensity stabilized by an acousto-optic modulator.  The transition of interest is driven by the lower sideband ($\approx 357.58\,$MHz) of a wideband electro-optic modulator (EOM) which is frequency shifted near to resonance for clock interrogation.  The clock laser propagates at an angle of $\theta = 3.0(1.5) ^\circ$ with respect to $\vop{x}$ (\fref{levels}e) and is linearly polarized in the plane spanned by the propagation vector and the magnetic field.  Control of the rf power driving the EOM enables equal $\pi$-times for each transition while maintaining same total laser intensity at the ion. 

A typical experiment consists of four steps: $500\,\mu\mathrm{s}$ of Doppler cooling, optical pumping for $10\,\mathrm{\mu s}$ to either $\ket{S_{1/2},m=\pm 1/2}$, a 600 $\mu$s clock interrogation pulse on a $\ket{D_{5/2},m^\prime}$ transition, and finally detection for $1\mathrm{ms}$.  The initial Doppler cooling step includes the 614-nm beam to facilitate repumping from the upper clock state. Second-order integrating servos~\cite{peik2005laser} independently track the average and difference frequency~\cite{bernard1998laser} for three pairs of Zeeman transitions $m=\pm\frac{1}{2} \rightarrow \pm m^\prime$ for $m^\prime = \{\frac{1}{2}, \frac{3}{2},\frac{5}{2}\}$~\cite{dube2013evaluation}. With a servo update period of $\approx0.8\,$s and probe duty cycle of 30$\%$, each Zeeman pair is servoed for approximately 1.5 minutes before cycling to the next.  Coherence times are limited by the background magnetic field noise such that the resonant transfer probabilities for $|m^\prime|=\{\frac{1}{2},\frac{3}{2},\frac{5}{2}\}$ transitions are approximately $\sim\{0.98,0.93,0.80\}$ for the 600 $\mu$s probe time.

\begin{figure} 
\includegraphics[width=\columnwidth]{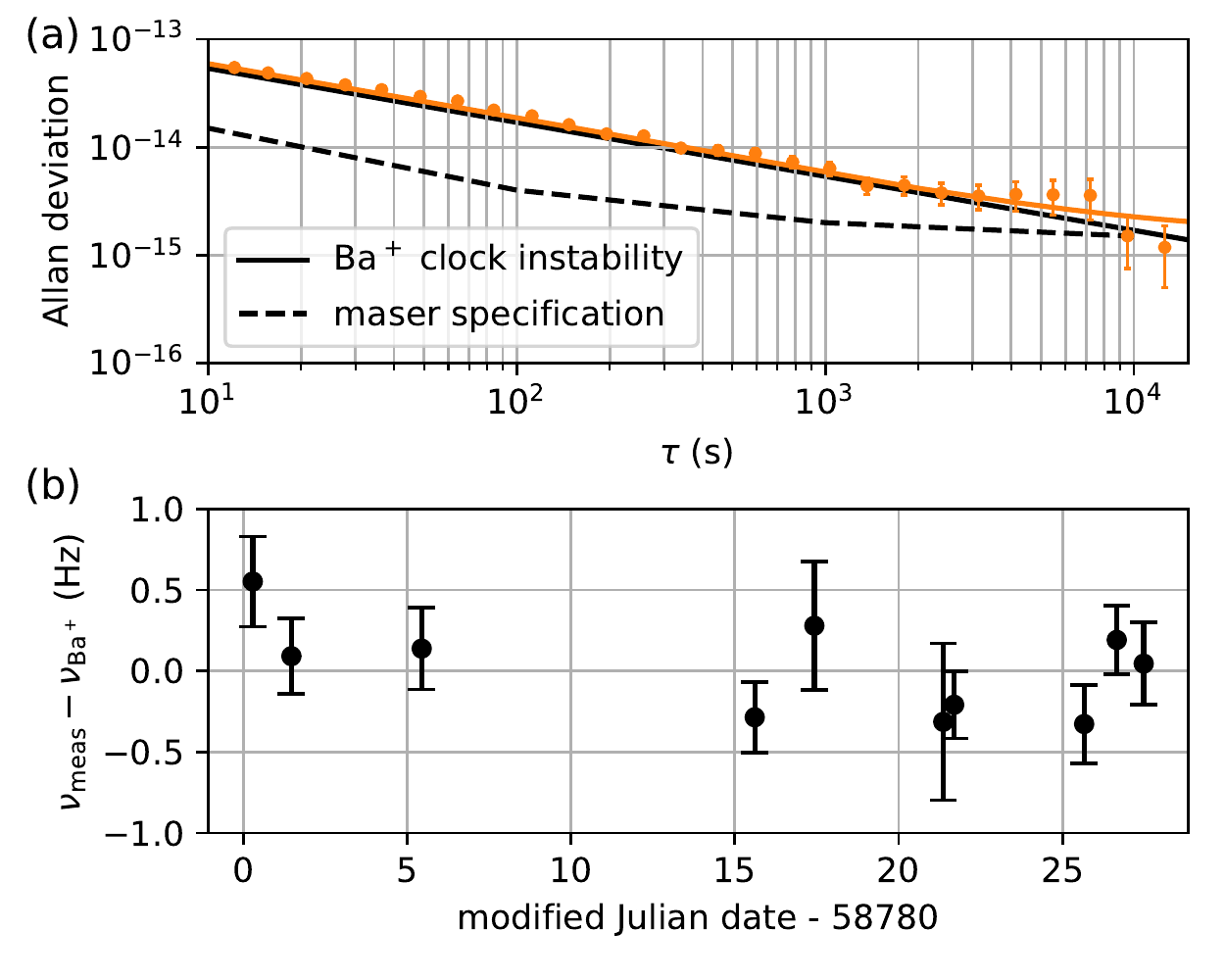}
\caption{(a) Allan deviation demonstrating the stability of the comparison between the barium clock and the HM (orange points). At shorter times the stability is consistent with projection noise limit of Ba$^+$ servo (black line), and at longer times approaches the $1.5\times10^{-15}$  frequency flicker noise floor of the HM (black dashed line). (b) Absolute frequency measurements given relative to $\nu_{\mathrm{Ba}^+}=170\,126\,432\,449\,333.31\,$Hz.}
\label{clockresult}
\end{figure}

The absolute frequency of the $S_{1/2} - D_{5/2}$ transition is obtained from the average of all six transitions, which is free of linear Zeeman and tensor shifts~\cite{dube2005electric}.  A total of ten measurements were taken over one month, with total measurement time of 70 h. The fractional instability of a typical measurement is shown \fref{clockresult}a (orange points). The statistical uncertainty of individual measurements shown in  \fref{clockresult}b is taken from the quadrature sum (orange line) of the Ba$^+$ servo instability (solid black line) and HM instability (black dashed line). The HM frequency is evaluated using GPS comparison data from a 20-day interval centered on the respective optical measurement, and with the assumption the HM has only a linear frequency drift over that time window. The total uncertainty in the HM calibration, including the link to the SI second via the Circular T, is evaluated to be $2.3 \times 10^{-15}$~\cite{SM}. Given the substantial overlap of the 20-day averaging windows used, the uncertainty in the maser calibration is not averaged down but taken as the statistical uncertainty in the final value. The mean result of all optical frequency measurements is $\nu_{\mathrm{Ba}^+}=170\,126\,432\,449\,333.31\pm(0.39)_\mathrm{stat}\pm(0.29)_\mathrm{sys}\,$Hz, including correction for the systematic effects listed in  \tref{tab:sys}.

\begin{table}
\caption{Systematic frequency shifts and uncertainties for the $^{138}$Ba$^+$ $S_{1/2} - D_{5/2}$ transition.}
\label{tab:sys}
\begin{ruledtabular}
\begin{tabular}{l c c}
Description & $\mathrm{shift} (\mathrm{Hz})$ & $\sigma$ (Hz) \\\hline
\midrule
black body radiation & 0.66 & 0.09\\
ac Stark (1762 nm)  &  0.53&0.27\\
quadratic Zeeman & 0.05&$<$0.01 \\
gravitational redshift & 0.32 & 0.06\\ \hline
total&1.56&0.29\\ 
\end{tabular}
\end{ruledtabular}
\end{table}

The leading systematic shifts and uncertainties for the $^{138}$Ba$^+$ clock transition are given in \tref{tab:sys} and each briefly discussed in the following paragraphs.  Averaging over all $m^\prime$ states eliminates tensor shifts which include the electric quadrupole shift, and tensor components of the quadratic Zeeman shift and ac-Stark shift from the 1762-nm laser.  Other systematic effects which are considered, but omitted from \tref{tab:sys} because they are well below the final uncertainty, include: shifts arising from excess micromotion (EMM), ac-Stark shifts due to the leakage of 493-nm or 615-nm laser light, second-order Doppler shifts from the thermal motion of the ion, and the ac-Stark shift due to off resonant quadrupole couplings of the clock laser. For completeness, a discussion of these shifts is given in the Supplemental Material. 

As reported in the Ref.~\cite{chanu2019magic}, the static and dynamic differential scalar polarizabilities are evaluated to be $\Delta \alpha_0(0) = -73.56\,(42)\,$a.u. and $\Delta \alpha_0(\omega_{0}) = -78.51\,(42)\,$a.u., where a.u. denotes atomic units, $\omega_{0}$ is the clock transition frequency, and we have taken the $2\sigma$ uncertainties as recommended in Ref.~\cite{chanu2019magic}.  The blackbody radiation (BBR) shift is evaluated for a temperature at ion of $T = 303\,(10)\,$K, 5 degrees above the ambient temperature in the laboratory. 

For the ac Stark shift due to the 1762-nm clock laser, the laser intensity is inferred from the observed Rabi coupling rates, accounting for the laser polarization, incidence angle relative to the magnetic field, $\phi-\theta=30(2)^\circ$, and the reduced electric-quadrupole matrix element $\bra{D_{5/2}}|r^2|\ket{S_{1/2}} = 15.80\,$a.u.~\cite{iskrenova2008theoretical}.  The EOM modulation depth is set to maximize the first-order sideband when driving the weakest transitions, $m=\pm\frac{1}{2}\rightarrow m^\prime=\pm\frac{5}{2}$.  The modulation depth is reduced accordingly to give equal coupling on the other transitions for constant intensity on ion. 

The quadratic Zeeman shift arises from Zeeman coupling between $D_{5/2}$ and $D_{3/2}$ fine structure states. The magnetic dipole matrix element is calculated, in the LS-coupling regime, to be $\bra{D_{5/2}}|M1|\ket{D_{3/2}} = 1.549\,(\mathrm{a.u.})$, which is consistent with other perturbative theory methods~\cite{safronova2011excitation} to better than 1\%. The quadratic Zeeman sensitivity for the average of all $| D_{5/2}, m^\prime\rangle $ states is evaluated to be $1.09(2)\,\mathrm{Hz}\,\mathrm{mT}^{-2}$ and all measurements were performed at a magnetic field near 0.225$\,\mathrm{mT}$. 

To determine the gravitational redshift we have taken a local geoid height of 7.9 m relative to the World Geodetic System (WGDS84) ellipsoid from the 2008 Earth Gravitation Model (EGM 2008) datum. The elevation of the ion in WGDS84 is determined from its relative position to the rooftop GPS receiver.  From this, we estimate the local height above the geoid to be 17(3) m. Given the gravitational acceleration of 9.780$\,\mathrm{m}\,\mathrm{s}^{-2}$ at equatorial latitude, the gravitational redshift is evaluated to be 0.32(6) Hz.

\begin{figure}
\includegraphics[width=\columnwidth]{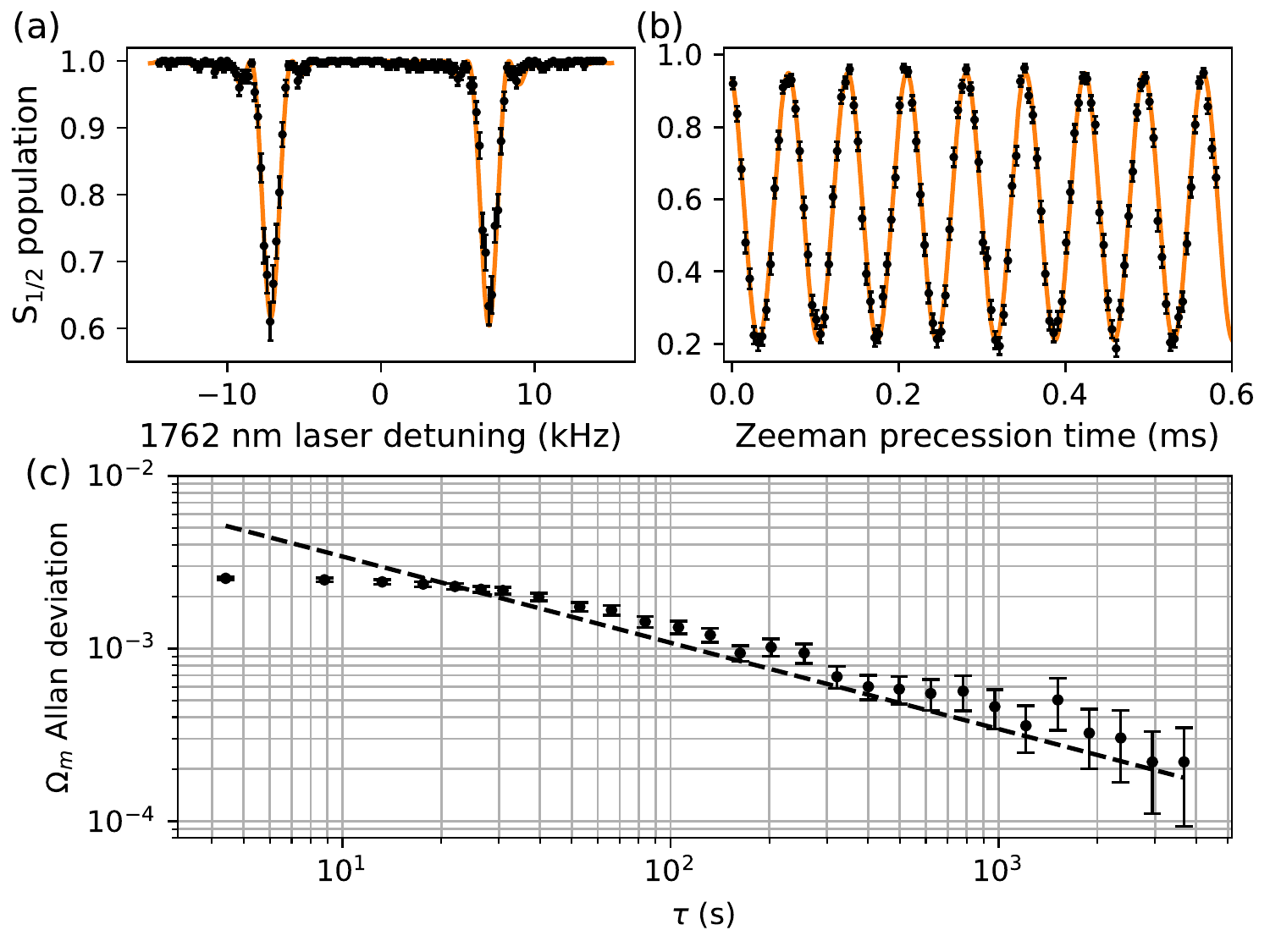}
\caption{(a) Autler-Townes splitting due to resonant ($\omega_S=\Omega_\mathrm{rf}$) transverse ac-magnetic field.   A fit (orange line) is obtained by $\chi^2$ minimization of the exact Hamiltonian solution from which we find the coupling strength is $\Omega_\mathrm{m}/2\pi=14.10(4)\,\mathrm{kHz}$ with reduced $\chi^2=0.94$. (b) When $\Omega_\mathrm{o}\gtrsim \Omega_\mathrm{m}$, Zeeman precession is observed. A fit (orange line) to a cosine function yields a precession frequency of $14.05(1)\,\mathrm{kHz}$. (c) Allan deviation for the measured Autler-Townes splitting.}
\label{ATsplitting}
\end{figure}

For determination of $g_D$, four transitions are used: $\ket{S_{1/2},m=\pm 1/2}\equiv\ket{S,\pm}$ to $\ket{D_{5/2},m=\pm 1/2}\equiv\ket{D,\pm}$, and $\ket{S,\pm}\leftrightarrow\ket{D,\mp}$ (see \fref{levels}b). The following differences in the $g_D$ experiments are noted: $\theta = -18^\circ$, $\phi = 0$, the clock probe times used were between 0.3-0.4 ms, the clock laser EOM is operated near the rf frequency $\approx167.55\,$MHz, and the 1762 nm laser intensity was not actively stabilized with the AOM which was instead used to switch the intensity during clock interrogation.  

In a static magnetic field, $B_0$, the Zeeman splittings for the $S_{1/2}$ and $D_{5/2}$ levels are given by $\omega_{S}=g_{S} \mu_B B_0/\hbar$ and $\omega_{D}=g_{D} \mu_B B_0/\hbar$ respectively.  The ratio $r=\omega_D/\omega_S=g_D/g_S$ can be inferred from measured frequency differences between the four transitions $\ket{S,\pm}\leftrightarrow\ket{D,\pm}$ and $\ket{S,\pm}\leftrightarrow\ket{D,\mp}$ from which $g_D=r g_S$ can be obtained using $g_S=2.002\,491\,92(3)$ given in \cite{marx1998precise}.  However, an rf-magnetic field, which arises from currents in the electrodes driven by the rf-trapping potentials, results in a measured ratio, $\tilde{r}$, given by~\cite{gan2018oscillating}
\begin{equation}
\label{eq:aczeemanshift}
\tilde{r}=\left[\frac{1+\frac{1}{2}\frac{\omega_D^2}{\omega_D^2-\Omega_\mathrm{rf}^2}\frac{\langle B_\perp^2 \rangle}{B_0^2}}{1+\frac{1}{2}\frac{\omega_S^2}{\omega_S^2-\Omega_\mathrm{rf}^2}\frac{\langle B_\perp^2 \rangle}{B_0^2}}\right]r
\end{equation}
where $B_\perp$ is the amplitude of the rf-magnetic field perpendicular to the applied static field.  For a given $B_\perp$, the correction factor in square parentheses can, to a good approximation, be inferred from the measured values of $\omega_S$, and $\omega_D,$ using the measured $g_S$ to determine $B_0$.

An accurate determination of $B_\perp$ can be obtained by setting $\omega_S=\Omega_\mathrm{rf}$, and driving the optical transition $\ket{S,+}\leftrightarrow\ket{D,+}$ with a coupling strength $\Omega_\mathrm{o}$.  When $\Omega_\mathrm{o}\ll\Omega_\mathrm{m}=g_S \mu_B B_\perp/(2\hbar)$, an Autler-Townes splitting on the optical transition is observed, as illustrated in \fref{ATsplitting}(a), with the two peaks separated by $\Omega_\mathrm{m}$.  When $\Omega_\mathrm{o} \gtrsim \Omega_\mathrm{m}$, precession of the ground-state population is observed, as illustrated in \fref{ATsplitting}(b), with the oscillation frequency given by $\Omega_\mathrm{m}$.  Both approaches give good agreement, with the data from \fref{ATsplitting} giving an inferred coupling $\Omega_\mathrm{m}/(2\pi)$ of $14.10(4)\,\mathrm{kHz}$ from the Autler-Townes splitting and $14.05(1)\,\mathrm{kHz}$ from the Zeeman precession frequency.  Although these are in agreement, an Autler-Townes splitting is a better method as an off-resonant Zeeman coupling appears as an imbalance in the two peaks and can be accounted for in data analysis.

We test the stability of $\Omega_\mathrm{m}$ by tracking the Autler-Townes splitting with a servo. The half-width half-maximum (HWHM) of the both lines in the Autler-Townes feature (such as shown in \fref{ATsplitting}(a)) are sequentially interrogated on both $\Delta m=0$ transitions, $\ket{S,\pm}\leftrightarrow\ket{D,\pm}$.  An outer servo loop adjusts the magnetic field to maintain the condition $\omega_S=\Omega_\mathrm{rf}$.  Measured continuously for over 4 hours, the observed splitting is projection noise limited, as illustrated in \fref{ATsplitting}(c), and shows a fractional stability of $\lesssim2\times10^{-4}$ at one hour. 

\begin{figure}
\includegraphics[width=\columnwidth]{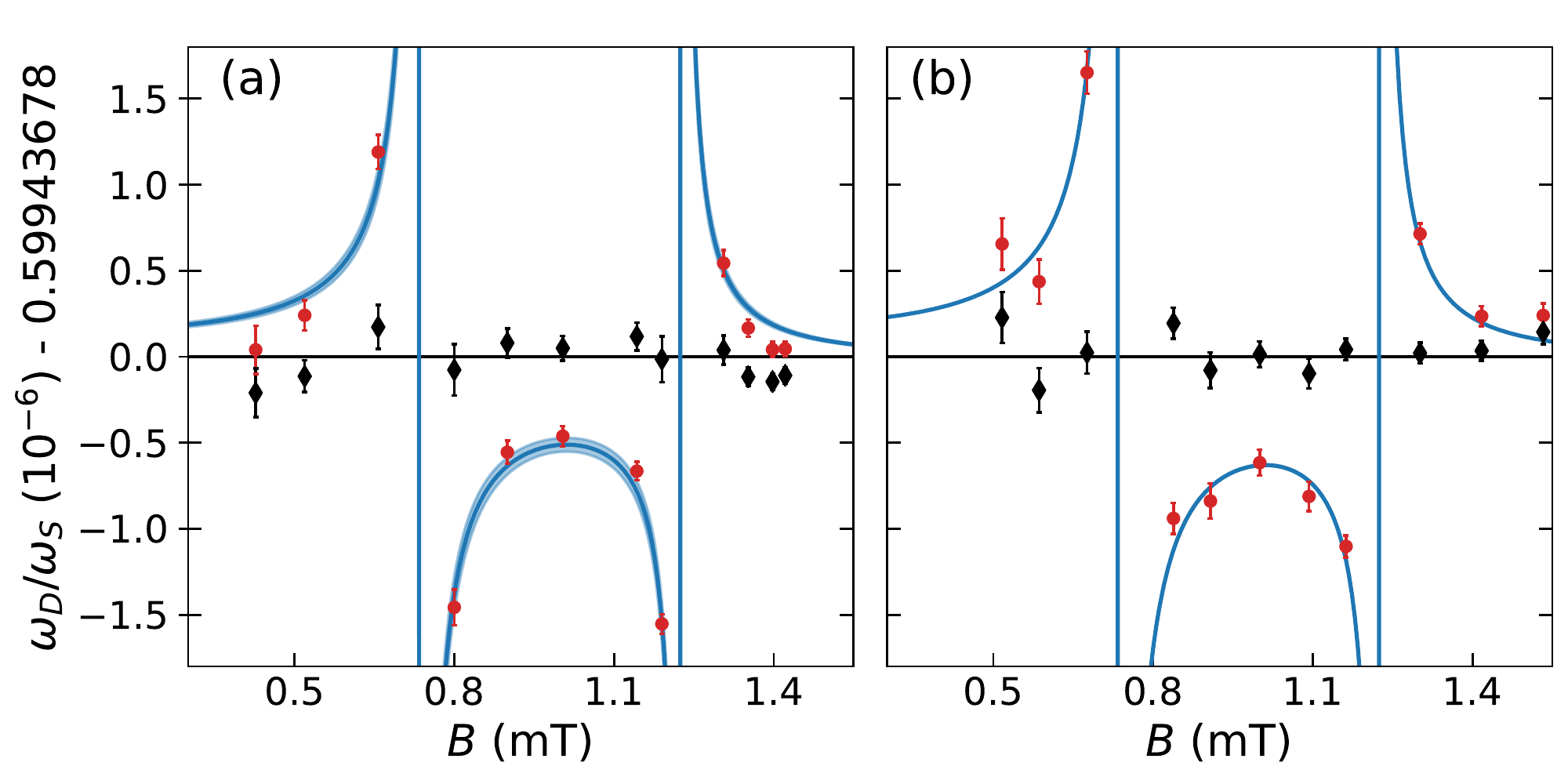}
\caption{(a,b) Two data sets measuring the ratio $\omega_D/\omega_S$ over a range of magnetic fields. The red circles are the uncorrected measurements of the ratio ($\tilde{r}$).  Black diamonds are ratios ($r$) corrected for the rf-magnetic field effects.  The blue line (shaded region) indicates the correction (uncertainty) evaluated by \eref{eq:aczeemanshift} given the measured $B_\perp = 0.906(35)\,\mu$T and $1.006(3)\,\mu$T for (a) and (b) respectively.}
\label{gJMeasurement}
\end{figure}

Measurements of $\omega_D/\omega_S$ were taken over a range of magnetic fields spanning $\sim 0.4-1.6\,\mathrm{mT}$, and two data sets taken approximately 1 month apart are shown in \fref{gJMeasurement}.   For each set, solid circles are the measured values  of $\omega_D/\omega_S$ with the magnetic field deduced from the measured ground-state splitting and the previously specified value of $g_S$; diamonds are corrected using values of $\Omega_\mathrm{rf}$ and $B_\perp$ measured at the time the data was taken.  Each point is a result of servoing on the four transitions for approximately $10^3$ servo updates with one servo update derived from $100$ measurements on either side of each of the transitions.   The error bars for the uncorrected data are the statistical errors from the servo, and the corrected points include the error arising from the uncertainty in $B_\perp$ as specified in the caption.

Using a $\chi^2$-minimization to determine the mean, the data in \fref{gJMeasurement}(a) gives a reduced $\chi^2$ of 1.95 indicating that the measurements are not limited by the statistical errors.  At that time, it was noted that the rf-resonator was not optimally coupled, which resulted in a degraded stability of the rf-confinement and hence $B_\perp$.  This was fixed for the second data set, \fref{gJMeasurement}(b), which gives which gives a statistically acceptable reduced $\chi^2$ of 1.4.  Nevertheless we take as the ratio result as the mean of the second data set with the full standard deviation as the uncertainty, $0.599\,436\,81(12)$, noting that this is also consistent with $0.599\,436\,72(11)$ from the first dataset. The ratio combined with $g_S = 2.002\,491\,92(3)$~\cite{marx1998precise} yields $g_D=1.200\,367\,39(24)$.

In summary, we have carried out precision spectroscopy of the $S_{1/2} - D_{5/2}$ clock transition in $^{138}$Ba$^+$.  The measurements have provided an absolute frequency determination of the clock transition with sub-Hertz level accuracy and a 30-fold improvement in the Land\'{e} $g_J$-factor for the $D_{5/2}$ level.  To our knowledge this is the first direct measurement of the optical transition frequency. A recently reported measurement of 146\,114\,384.0(1)\,MHz for the $S_{1/2} - D_{3/2}$ transition~\cite{dijck2015determination}, together with the fine structure splitting of 24\,012\,048\,319(1)\,kHz~\cite{whitford1994absolute}, is consistent with our result albeit limited by the 100\,kHz uncertainty of ~\cite{dijck2015determination}. Our value for $g_D$ is in agreement with the value reported in \cite{hoffman2013radio}. The discrepancy with the result of  \cite{lewty2013experimental} is potentially due to that quadrupole shifts, which were not considered in that anlysis.  We have also demonstrated the influence that ac-magnetic fields arising from rf-currents in the electrodes can have on $g_J$-factor measurements.  Similar to the microwave demonstration reported in \cite{gan2018oscillating}, the observation of an Autler-Townes splitting on the clock transition has allowed an accurate characterization of the ac-magnetic field component orthogonal to the applied dc field.  This will be an important diagnostic tool for characterizing the ac-magnetic field shift for the Lu$^+$ clock transitions \cite{gan2018oscillating}.

\begin{acknowledgements}
This work is supported by the National Research Foundation, Prime Ministers Office, Singapore and the Ministry of Education, Singapore under the Research Centres of Excellence programme. This work is also supported by A*STAR SERC 2015 Public Sector Research Funding (PSF) Grant (SERC Project No: 1521200080).
\end{acknowledgements}

\bibliographystyle{apsrev4-1}
\bibliography{BaClock}

\pagebreak



\section*{Supplementary material}

# Supplemental Material for Precision measurements on the $^{138}$Ba$^+$ $6s\ ^2S_{1/2} - 5d\ ^2D_{5/2}$ clock transition

K. J. Arnold,[1,2,*] R. Kaewuam,[1] S. R. Chanu,[1] T. R. Tan,[1,3] Zhiqiang Zhang,[1] and M. D. Barrett[1,3,†]

[1]*Centre for Quantum Technologies, National University of Singapore, 3 Science Drive 2, 117543 Singapore*
[2]*Temasek Laboratories, National University of Singapore, 5A Engineering Drive 1, 117411 Singapore*
[3]*Department of Physics, National University of Singapore, 2 Science Drive 3, 117551 Singapore*
(Dated: March 12, 2020)

Supplemental information on the calibration of the hydrogen maser via GPS comparison, and discussion of other $S_{1/2} - D_{5/2}$ clock systematics not detailed in the main article which contribute negligibly ($< 10^{-16}$ fractional frequency uncertainty).

## I. MASER CALIBRATION

The absolute frequency calibration of the active hydrogen maser (Microsemi MHM 2010) is assessed by long-term logging of the pulse-per-second (PPS) time difference relative to a commercial GPS-synchronized timebase (Meridian II from Endrun Technologies Inc). The Meridian II is a single band (L1) GPS receiver design which employs a proprietary algorithm for real-time ionospheric correction based on L1 channel code and carrier phase divergence. Fig. 1a (blue points) shows the measured time difference, averaged every $10^3$ s,

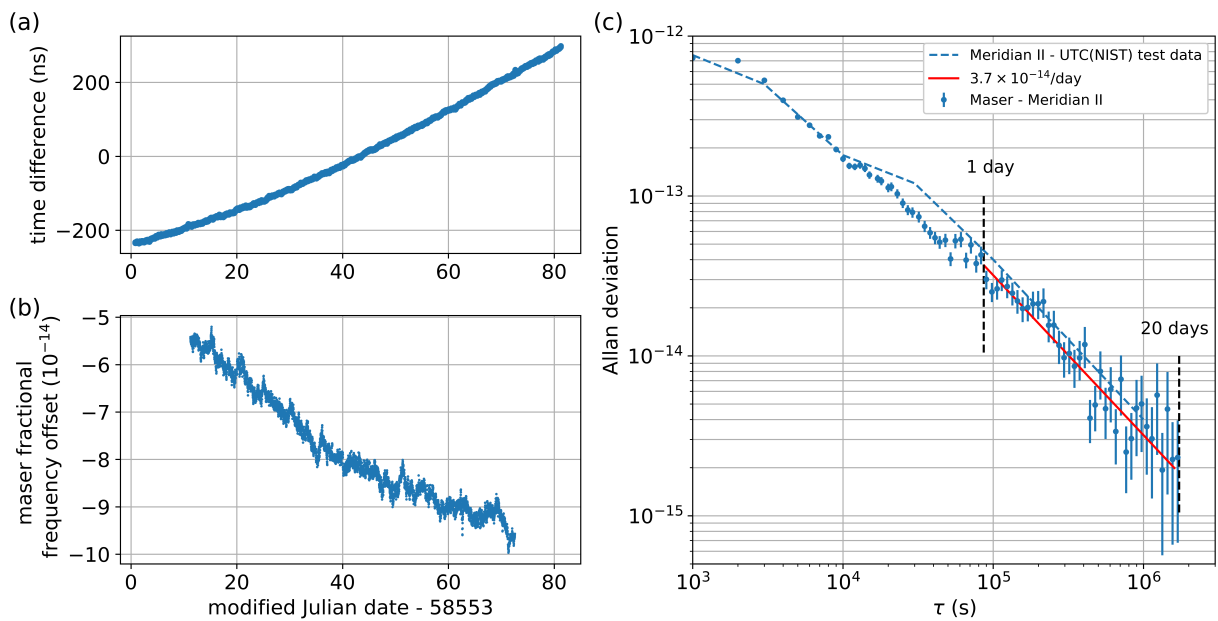

FIG. 1. (a) Measured PPS time difference between the Meridian II GPS timebase and maser spanning over 80 days. (b) Maser fractional frequency offset relative to GPS time assessed using a 20-day averaging window centered at each point. (c) Allan deviation (blue points) of Meridian II GPS timebase compared to the maser after subtracting the maser frequency offset shown in (b). The instability is consistent with the test data for a Meridian II model which was compared to UTC(NIST) and reported by the manufacturer (blue dashed). The fitted asymptote of $3.7(2) \times 10^{-14}$/day (red line) is taken as the instability of the maser-GPS link for averaging times between 1 and 20 days.

---

* cqtkja@nus.edu.sg
† phybmd@nus.edu.sg

between the maser and Meridian II for the longest uninterrupted logging interval available which we use here for estimating the GPS link stability. A feedforward applied at a constant rate compensates a $-3.6 \times 10^{-15}$ linear frequency drift of the maser, which was the frequency drift rate assessed several months prior. The maser frequency relative to the Meridian II is assessed at each point from a centered 20-day window and fit assuming only a linear frequency drift and offset for the maser. Fig. 1b shows the fractional frequency offset of the maser assessed by this method.

Fig. 1c (blue points) shows the instability of the Meridian II as compared against the hydrogen maser after subtracting the maser frequency offset, which is shown Fig. 1b. The instability is consistent with the test data for a Meridian II model taken at NIST (blue dashed line) which is supplied by the manufacturer. The fitted asymptote (red line in Fig. 1c) is used to estimate the statistical uncertainty of maser frequency due to the GPS link stability. For the 20 day averaging window used, the fractional frequency uncertainty is $1.9 \times 10^{-15}$.

The maser frequency assessment against GPS time is linked to the SI second by the monthly circular T reports. Section 4 of the circular T publishes the daily frequency time difference between GPS time and UTC obtained using the values [UTC-UTC(OP)] and the GPS data aquired at the Paris Observatory (OP). Inclusion of these time differences shifts the $\nu_{\text{Ba}^+}$ result by $0.2(1.3) \times 10^{-15}$ fractionally. The uncertainty of $1.3 \times 10^{-15}$ is estimated from the 20-day instability of the UTC-GPS data in Section 4 of the circular T over the previous two years. The evaluation against UTC, or equivalently international atomic time (TAI) which is offset from UTC by an integer number of seconds, is finally linked to the SI second by the fractional frequency deviation of TAI estimated in section 3 of the Circular T. This shifts the $\nu_{\text{Ba}^+}$ result by $3.3(1.4) \times 10^{-16}$ fractionally. The total uncertainty of the maser calibration to the SI second is estimated to be $2.3 \times 10^{-15}$ found by the quadrature sum of the GPS-link instability at 20-days, $1.9 \times 10^{-15}$, the UTC-GPS instability at 20-days, $1.3 \times 10^{-15}$, and the uncertainty in the UTC-SI correction reported in the circular T, $1.4 \times 10^{-16}$.

## II. OTHER $S_{1/2} - D_{5/2}$ CLOCK SYSTEMATICS

Estimations of other possible systematic effects which have been considered, but not included in Table I of the main article, are briefly summarized.

**Excess micromotion (EMM):** EMM is compensated periodically in all three directions using a combination of two methods: sideband spectroscopy on the 1762-nm clock transition and photon-correlation of the fluorescence collected while Doppler cooling [1]. Total EEM shifts are estimated to be less than $10^{-17}$ fractionally.

**493-nm leakage light:** All 493-nm laser paths are switched by two cascaded double-pass acousto-optic modulators, with a combined extinction better than -140 dB, eliminating any possibly of significant leakage light.

**614-nm leakage light:** For the 614-nm repump laser, the ac-Stark is also negligible for the minimal intensity used and measured switching extinction of -75 dB.

**Second-order Doppler:** The ion temperature is inferred from the thermal dephasing when Rabi flopping on the $|S_{1/2}, m = 1/2\rangle \leftrightarrow |D_{5/2}, m' = 1/2\rangle$ carrier transition as shown in Fig. 2. The fit model [2] implies a temperature of $T = 2.4 T_\text{D}$ for the typical data shown in Fig. 2, where $T_\text{D} \sim 0.5 \,\text{mK}$ is the Doppler cooling limit. Here all dephasing is attributed to the thermal effect and thus is only an upper bound on the temperature. For an ion temperature $T \lesssim 3T_\text{D}$, the second-order Doppler shift due the thermal motion is then $\frac{\delta \nu_{\text{dopp}}}{\nu_{\text{Ba}^+}} = -\frac{3}{2} \frac{k_B T}{mc^2} \sim -1 \times 10^{-18}$.

**Quadrupole couplings:** Off-resonant quadrupole couplings of the clock laser to all possible Zeeman components gives rise to an ac Stark shift. By symmetry, this shift cancels in the average frequency of a pair of symmetric Zeeman transitions if the clock laser has a single frequency component, as noted in [3]. Here, we must additionally consider the couplings of all the clock laser frequency components generated by the EOM, which do not in general cancel in the average frequency. The EOM modulations depth of $\sim (0.65, 1.5, 1.9)$ were used for the $m = \pm\frac{1}{2}$ to $m' = (\pm\frac{1}{2}, \pm\frac{3}{2}, \pm\frac{5}{2})$ lines to achieve equal couplings on the respective transitions for our given clock laser geometry. Given the $\pi$-time of 0.6 ms

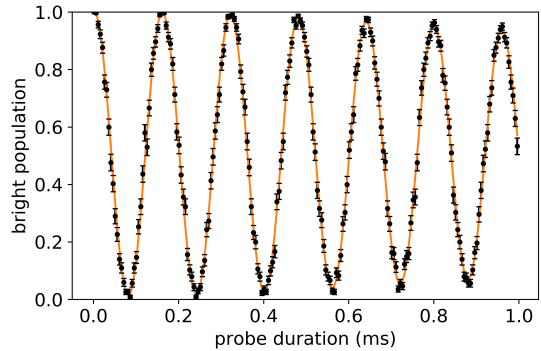

FIG. 2. Typical Rabi flopping on the 1762-nm optical clock transition demonstrating full contrast with thermal dephasing.

and adding the contributions from all off-resonant couplings, including sidebands out to order 100, the total shift in the average of all six transitions is estimated to be $< 0.1\,\mathrm{mHz}$, or $< 1 \times 10^{-17}$ fractionally.



\end{document}